\documentclass[aps,prb,twocolumn,superscriptaddress, show pacs]{revtex4-2}
\usepackage{bm, amsmath}
\usepackage{graphicx}
\usepackage{color}
\usepackage{epstopdf}
\usepackage{graphicx}
\usepackage{dcolumn}
\usepackage{bm}
\usepackage{subfigure}
\usepackage[rawfloats=true]{floatrow} 
\begin{document}

\title{Meron-mediated phase transitions in quasi-two-dimensional chiral magnets with easy-plane anisotropy: successive transformation of the hexagonal skyrmion lattice into the square lattice and into the tilted FM state}

\author{Andrey O. Leonov}
\thanks{Corresponding author: leonov@hiroshima-u.ac.jp}
\affiliation{Department of Chemistry, Faculty of Science, Hiroshima University Kagamiyama, Higashi Hiroshima, Hiroshima 739-8526, Japan}
\affiliation{International Institute for Sustainability with Knotted Chiral Meta Matter, Kagamiyama, Higashi Hiroshima, Hiroshima 739-8526, Japan}

\date{\today}

\begin{abstract}
{I revisit the well-known structural transition between hexagonal and square skyrmion lattices induced by increasing easy-plane anisotropy 
in quasi-two-dimensional chiral magnets. 
I show that the hexagonal skyrmion order, by the first-order phase transition, transforms into a distorted (rhombic) skyrmion lattice. 
The transition is mediated by merons and anti-merons emerging within the boundaries between skyrmion cells. Per hexagonal unit cell, there are two merons with the positive topological charge density (located in the corners of hexagons and shared by three neighbors) and three anti-merons with the negative topological charges (shared by two neighboring cells).  
Since the energy density associated with anti-merons is highly positive owing to the wrong rotational sense, 
one anti-meron per unit cell annihilates: 
anti-merons are squeezed by the pairs of approaching merons at the opposite sides of the hexagonal unit cell.
%
Further, in a narrow range of anisotropy values, the distorted skyrmion lattice gradually transforms into a perfect square order of skyrmions (alternatively called "a square meron-antimeron crystal") when two merons eventually merge into one.  Thus, within the square skyrmion lattice, there is one meron and two anti-merons per unit cell residing in the cell boundaries, which 
underlie the subsequent first-order phase transition into the tilted ferromagnetic state. A pair of oppositely charged merons mutually annihilates, whereas a remaining anti-meron couples with an anti-meron  occupying the center of the unit cell. Since two anti-merons have the opposite polarity, they form a bimeron, which perfectly fits into the homogeneous state. As an outcome, the tilted ferromagnetic state contains bimeron clusters (chains) with the attracting inter-soliton potential, just like the field-polarized state would accommodate isolated axisymmetric skyrmions dispersed after the skyrmion lattice expansion in the strong magnetic field. The findings of the paper shed new light on the role of merons as drivers of phase transitions between different states in chiral magnets. Moreover, domain-wall merons are actively involved in dynamic responses of the square skyrmion lattices.
As an example, I theoretically study spin-wave modes and their excitations by ac magnetic fields. Two found resonance peaks are the result of 
the complex dynamics of domain-wall merons: whereas in the high-frequency mode,  merons rotate counterclockwise as one might expect, in the low-frequency mode, merons are created and annihilated consistently with the rotational motion of the domain boundaries. 
}
\end{abstract}

\pacs{
75.30.Kz, 
12.39.Dc, 
75.70.-i.
}
         
\maketitle

\section{Introduction}

In magnetic compounds with broken inversion symmetry, the chiral crystal lattice induces a specific asymmetric exchange coupling, the so-called Dzyaloshinskii-Moriya interaction (DMI) \cite{Dz64,Moriya}. 
Within a continuum approximation for magnetic properties, the DMI is expressed by Lifshitz invariants (LI) -- the energy terms involving first derivatives of the
magnetization  $\textbf{m}$ with respect to the spatial coordinates $x_k$:
\begin{equation}
\mathcal{L}^{(k)}_{i,j} = m_i \partial m_j/\partial x_k - m_j  \partial m_i/\partial x_k
\label{Lifshitz}
\end{equation}
%
%
Depending on the crystal symmetry, certain combinations of the Lifshitz invariants can contribute to the magnetic energy of the material \cite{JETP89,Le}.

In a general case of cubic helimagnets, such as the itinerant magnets, MnSi \cite{Kadowaki,Muehlbauer09} and FeGe \cite{FeGe}, and the Mott insulator, Cu${_2}$OSeO${_3}$ \cite{Seki2012,Weiler}, the DMI reduces to the following concise form \cite{Dz64,JETP89}:
\begin{align}
w_D= \mathcal{L}^{(x)}_{z,y} + \mathcal{L}^{(y)}_{x,z} + \mathcal{L}^{(z)}_{y,x} = \mathbf{m}\cdot \mathrm{rot} \mathbf{m}.
\label{DMI}
\end{align}

In polar magnets with the C$_{nv}$ symmetry, such as  GaV$_4$S$_8$  and GaV$_4$Se$_8$ \cite{Bordacs17,Fujima17}, the DMI energy density,  
\begin{equation}
w_D=  m_x\partial_x m_z-m_z\partial_x m_x+m_y\partial_y m_z-m_z\partial_y m_y,
\label{Cnv}
\end{equation}
does not include LIs along the high-symmetry $z$ axis.

DMI of the same functional form (\ref{Cnv}) is also induced in multilayered structures due to the breaking of the inversion symmetry at interfaces, as occurs, e.g., in  PdFe/Ir (111) \cite{Romming2013}. Such artificial systems, enabled by the possibility of stacking, are extremely versatile as regards the choice of the magnetic, non-magnetic, and capping layers.

\begin{figure*}
\includegraphics[width=1.99\columnwidth]{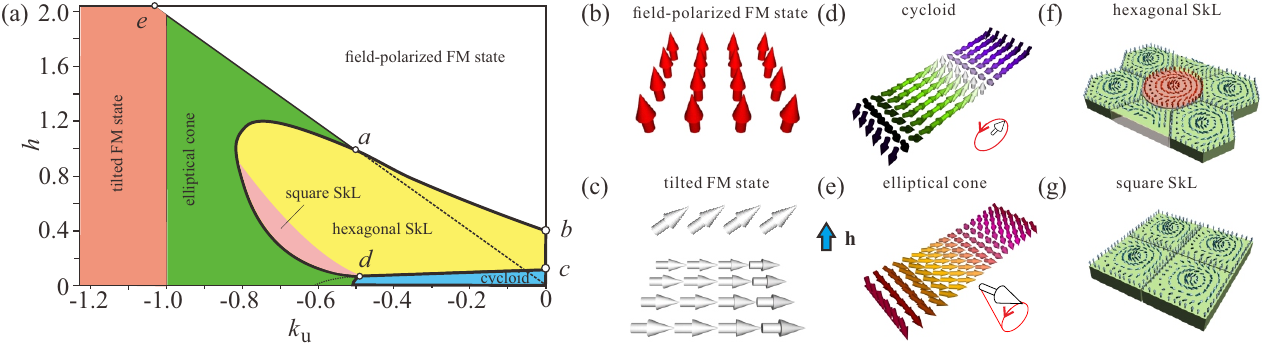}
\caption{(color online) (a) Magnetic phase diagram of the solutions for model (\ref{functional}) with the easy-plane uniaxial anisotropy.  Filled areas designate the regions of thermodynamical stability of corresponding phases: white shading - polarized ferromagnetic state (b); red shading - tilted ferromagnetic state (c); blue shading - cycloidal spiral (d), green shading - elliptical cone (e); yellow and pink shading - hexagonal (f) and square (g) skyrmion lattices. The field is measured in the units of $H_0 = D^2/A|\mathbf{M}|$, i.e., $\mathbf{h} = \mathbf{H}/H_0$. $k_u=K_uA/D^2$ is the non-dimensional anisotropy constant. In the following simulations, $h=0.5$ whereas the anisotropy constant is varied.
\label{fig01}}
\end{figure*}

LIs (\ref{Lifshitz})  are indispensable  to overcome the constraints of the Hobart-Derrick theorem \cite{Hobart,Derrick}, and thus to yield a set of competing modulated phases (generally, multidimensional) as well as countable solitons in the  phase diagrams specific to different crystallographic classes.

Since in cubic helimagnets, the rotational terms (\ref{DMI}) are present along all three spatial directions, the modulated phases and solitons are expected to be truly three-dimensional (3D). For example, magnetic hopfions are torus-shaped solitons embedded into a homogeneously magnetized background and characterized by the linked preimages \cite{hopfion,hopfion2}.  Generally, the magnetic phases in cubic chiral magnets develop additional twists involving all LIs near the surfaces (so-called "surface twists"), which are known to be essential for their thermodynamic stability \cite{Rybakov2013}. 
Recently, skyrmion lattice states (SkL) and isolated skyrmions (ISs) were discovered in bulk crystals of chiral magnets near the magnetic ordering temperatures \cite{Muehlbauer09,FeGe} and in nanostructures with confined geometries over larger temperature regions \cite{yuFeCoSi,yuFeGe}.

Skyrmions generate enormous interest due to the prospects of their applications in information storage and processing \cite{Sampaio13,Tomasello14,Shigenaga}.
Indeed, skyrmions are topologically protected \cite{Cortes-Ortuno}, they have the nanometer size \cite{Wiesendanger2016} and can be manipulated by electric currents \cite{Schulz,Jonientz}.  Skyrmions are also interesting objects for magnonics, e.g., collective spin dynamics within SkLs exhibits two spin-wave modes with the clock-wise (CW) and counterclockwise (CCW) rotation of skyrmions for the in-plane ac magnetic field as well as one breathing mode for the out-of-plane ac magnetic field \cite{Mochizuki}.

For the C$_{nv}$ symmetry, only modulated magnetic structures with wave vectors perpendicular to the polar axis are favored by the DMI (\ref{Cnv}) and thus represent 2D motifs of the magnetization. 
The phase diagrams constructed for such quasi-two-dimensional chiral magnets, nevertheless,  are far from being simple. Fig. \ref{fig01} (a) shows the well-known phase diagram for chiral magnets with the easy-plane anisotropy   (EPA) \cite{Lin,Tretiakov}. Besides homogeneous field-polarized  and tilted states (Fig. \ref{fig01} (b), (c)), the phase diagram features one-dimensional cycloids and elliptical cones (Fig. \ref{fig01} (d), (e)). Moreover, it implies that two types of skyrmion orderings -- the square and the hexagonal SkLs -- are stable even though the system does not have any anisotropy axis within the plane 
(Fig. \ref{fig01} (f), (g)).  Notice that at the phase diagram of chiral magnets with the DMI (\ref{DMI}) and the easy-plane uniaxial anisotropy, the conical phase with the wave vector along the field is the only stable modulated state \cite{Rowland}, which thus makes the phases in Fig. \ref{fig01} energetically less favorable.

Some phase transitions at the phase diagram in Fig. \ref{fig01} (a) are well-understood: (i) the first-order phase transition between the cycloid and the hexagonal SkL (line $d-c$ in Fig. \ref{fig01} (a)) occurs via ruptures of the the cycloidal state called meron pairs, which acquire the energetic advantage above this critical field (see, e.g.,  Refs. \cite{Mukai22,Muller,Ezawa} for details); (ii) the second-order phase transition between a skyrmion lattice and the field-polarized FM state (line $a-b$ in Fig. \ref{fig01} (a)) occurs via the infinite expansion of the lattice period \cite{Bogdanov94,Bogdanov99}; (iii) the elliptical cone and the tilted FM state gradually align along the field at the line $a-e$.  

Other phase transitions at the phase diagram are less understood. One can only anticipate that the anisotropy-driven phase transitions between hexagonal and square SkLs as well as between the square SkLs and 1D elliptical cones are of the first order \cite{Lin,Tretiakov} since they occur between topologically incompatible phases.

\begin{figure*}
\includegraphics[width=1.5\columnwidth]{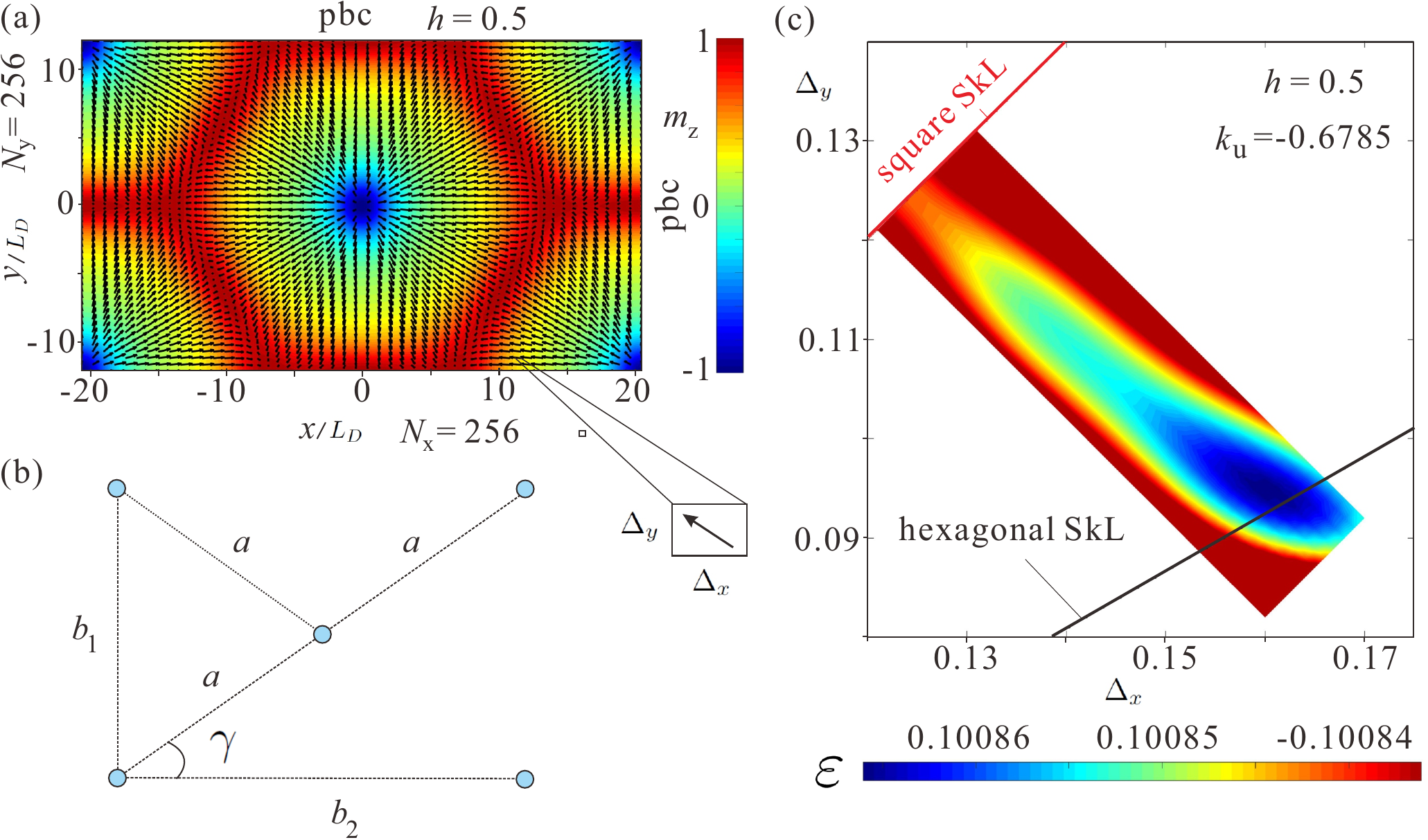}
\caption{(color online) (a) Schematics of a computational unit cell corresponding to a distorted (rhombic) SkL. The distribution of the magnetization within skyrmions retains its axisymmetric circular shape. The number of discretization points is equal along $x$ and $y$, $N_x=256, \, N_y=256$. The cell sizes, on the contrary, are varied to search for a deformed SkL with the lowest energy density.  (b) The characteristic geometric parameters of the unit cell, which exhibit the inter-skyrmion distances $b_1, b_2$, and $a$ as well as the characteristic angle $\gamma$. (c)  The energy density of distorted SkLs computed by integration of  (\ref{functional}) for different values of lattice spacings and for $h=0.5,\, k_u=0.6785$. The well-discernible energy minimum is formed for the skyrmion ordering, which is almost hexagonal.  The length scale is measured in units of $L_D$ (\ref{units}).
\label{fig02}}
\end{figure*}

In the present manuscript, I re-examine the mentioned unclear transitions. I show that the reorientation transition between the hexagonal and the square skyrmion arrangements with the increasing anisotropy value occurs via distorted (rhombic) SkLs (Sect. \ref{sect03}). The distorted SkLs represent the global minima of the system and gradually transit into the square SkL. The hexagonal SkL remains almost intact until its energy minimum disappears during this first-order phase transition.
I underline the decisive role of merons and anti-merons formed within the boundary regions between skyrmion cells in SkLs. Anti-merons with the negative topological charge density are shared by two adjacent skyrmion cells and bear the positive energy density. Since there are three such anti-merons within one hexagonal unit cell, the square cell with just two anti-merons becomes energetically more favorable even though the skyrmion packing density slightly decreases.  
During the structural transition, two corner merons with the positive topological charge density merge and thus annihilate one unfavorable anti-meron. Two such annihilation events at the opposite cell boundaries signify the reorientation transition from the hexagonal to the square skyrmion order. 

The phase transition between the square SkL and the tilted ferromagnetic state is also meron-mediated (Sect. \ref{sect04}). During this process, merons and anti-merons mutually collapse. Since the corner merons are shared by four neighboring skyrmion cells and anti-merons by two skyrmion cells, there are two anti-merons and one meron per each unit cell. After the collapse, the remaining anti-meron couples with the central anti-meron and forms a localized state known as a bimeron \cite{Mukai24}. Both anti-merons have the negative topological charge but the opposite polarity. 
As a result, the homogeneous state contains some finite number of bimerons (one per each unit cell) just like the field-polarized FM state would host isolated axisymmetric ISs. Hence, the findings of the present paper shed new light on the phase transitions among different phases and imply merons as important drivers guiding the whole process.

I also study collective spin dynamics of merons within the square skyrmion lattice (Sect. \ref{sect05}). I find two spin-wave resonances: (i) in the high-frequency mode,  the central anti-meron performs counter-clockwise rotation; (ii) in the low-frequency mode, the central anti-meron virtually does not move, but the domain boundary approaches it with each side sequentially; such a rotation of the domain-wall (DW) network is accompanied by creation and annihilation of DW merons.

\begin{figure*}
\includegraphics[width=1.99\columnwidth]{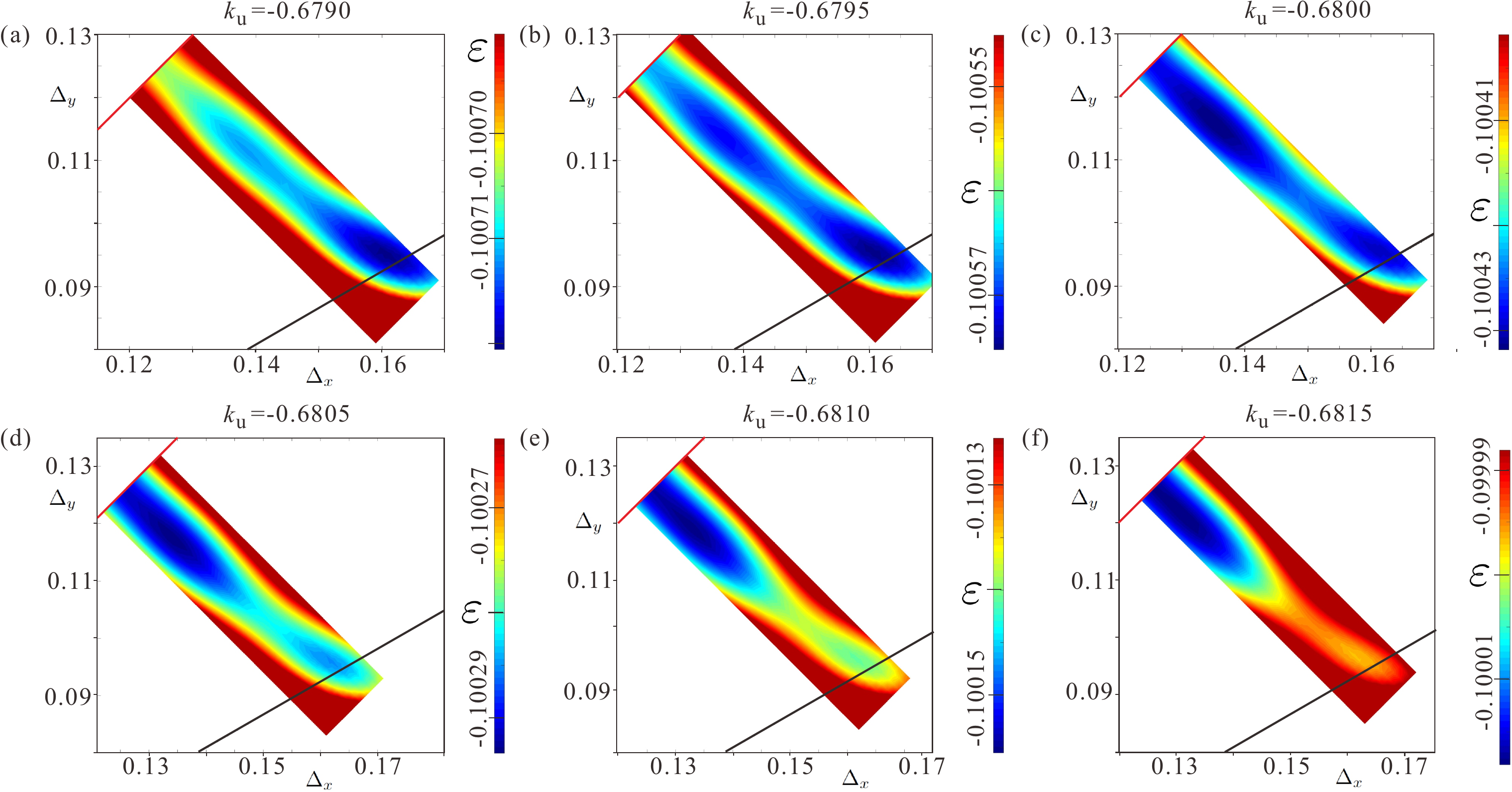}
\caption{(color online) The color plots of the energy densities $\varepsilon$ on the plane $\Delta_x$$\Delta_y$ for the increasing value of the uniaxial anisotropy $k_u$. Each energy plot features two energy minima: for low values of $k_u$, the global minimum corresponds to a hexagonal SkL; for larger $k_u$, the global minimum belongs to the distorted (rhombic) SkL on its way towards the square skyrmion order.
\label{fig03}}
\end{figure*}

\section{Phenomenological model \label{sect02}} 

The magnetic energy density of a two-dimensional noncentrosymmetric ferromagnet 
can be written as the sum of the exchange, the DMI (\ref{Cnv}), Zeeman, and the anisotropy energy contributions, correspondingly:
\begin{equation}
w(\mathbf{m})=\sum_{i,j}(\partial_i m_j)^2+w_{D}-\mathbf{m}\cdot\mathbf{h} - k_u m_z^2.
\label{functional}
\end{equation}
Here, we introduced the non-dimensional units  to make the results more general and 
to be directly mapped to any material system.  
Spatial coordinates $\mathbf{x}$ are measured in units of the characteristic length of modulated states $L_D$. 
$A>0$ is the exchange stiffness, $D$ is the Dzyaloshinskii constant. $k_u$ is the non-dimensional anisotropy constant, which leads to the easy-plane magnetization, i.e., $k_u<0$. 

\begin{align}
&L_D=A/D, k_u=K_uA/D^2, \nonumber\\
&\mathbf{h} = \mathbf{H}/H_0, H_0 = D^2/A|\mathbf{M}|.
\label{units}
\end{align}
$\mathbf{h}$ is the  magnetic field applied along $z$ axis.
The magnetization vector $\mathbf{m}(x,y)$ is normalized to unity. 
%

Alternatively, the length scale can be measured in units of $\lambda$:
\begin{equation}
\lambda=4\pi L_D,
\label{lambda}
\end{equation}
which is the period of the spiral state for zero magnetic field (e.g., 18 nm for the bulk MnSi or 60 nm for Cu$_2$OSeO$_3$ \cite{Seki2012,Crisanti}). 
In actual simulations, we measure the length in units of $L_D$ (\ref{units}). Dividing by $4 \pi$, we get the length scale in units of $\lambda$, which provides a direct comparison with a specific material system. We will use both length scales throughout the paper.

We consider a 2D film of a ferromagnetic material on the $xy$-plane using periodic boundary conditions (pbc). The value of the field is kept constant $h=0.5$ whereas the value of the anisotropy constant $k_u$ is changed to address the aforementioned phase transitions between modulated phases.

We neglect the influence of dipole-dipole interactions due to the magnetic charges  formed within different states with the Neel-like type of the magnetization rotation. We assume that the DM interactions suppress demagnetization effects and are the main driving force leading to the magnetization rotation and to the equilibrium periodicity. Moreover, the shape anisotropy in this case represents an additional correction of the easy-plane anisotropy. The influence of dipole-dipole interactions on the effects found in the present manuscript will be considered elsewhere.

As a primary numerical tool to minimize the functional (\ref{functional}), we use MuMax3 software package (version 3.10) which calculates magnetization dynamics by solving the Landau-Lifshitz-Gilbert (LLG) equation with finite difference discretization technique \cite{mumax3}.
To double-check the validity of obtained solutions, we also use our own numerical routines, which are explicitly described in, e.g., Ref. \cite{metamorphoses} and hence will be omitted here.

All structures are minimized on the grid $256\times 256 \times 1$. 
To check the stability of different skyrmion orderings, we compute the energy density (\ref{functional}) for different ratios of the grid spacings $\Delta_y$ and $\Delta_x$ (called cell sizes in mumax3, Fig. \ref{fig02} (a)). $\Delta_z=0.1$ remains the same in all simulations.
The axisymmetric distribution of the magnetization within skyrmion cores is preserved during this minimization procedure.  Thus, varying lattice spacings lead to the rearrangement of the constituent skyrmion cores spanning all possible lattice orders. 

\begin{figure*}
\includegraphics[width=1.5\columnwidth]{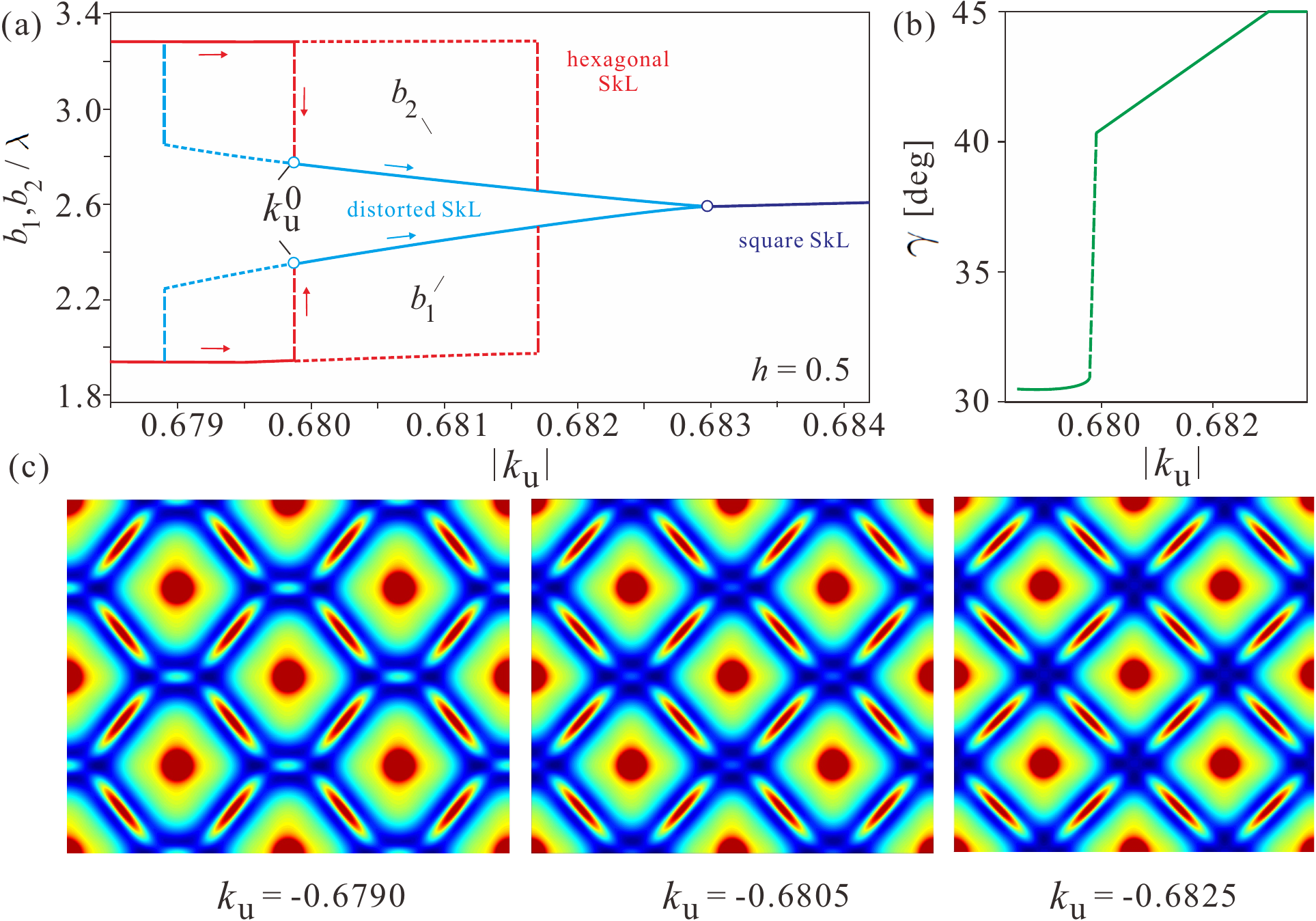}
\caption{(color online) (a) Equilibrium parameters of the unit cell, as introduced in Fig. \ref{fig02} (b), in dependence on the value of the EPA during the structural transition from the hexagonal into the square skyrmion arrangement. Red and blue lines correspond to the hexagonal and distorted SkLs, correspondingly, with the solid and dotted lines indicating the global and local energy minima. After some value of $k_u$, the square SkL (dark-blue line) is completely formed. 
(b) The gradual anisotropy-driven change of the angle $\gamma$ from the value $\gamma \approx 30^{\circ}$  in the hexagonal SkL to the value $\gamma \approx 45^{\circ}$  within the square SkL. (c) Color plots of the energy density $w(x,y)$ characterize the gradual evolution of the distorted SkL into a square one. 
\label{fig04}}
\end{figure*}

Fig. \ref{fig02} (a) shows the centered rectangular unit cell used for computations of skyrmion orderings.  
To characterize the degree of SkL deformations, we introduce the following lengths and angles consistent with the square and the hexagonal SkLs (blue circles correspond to skyrmion centers, Fig. \ref{fig02} (b)): (i) within the square SkL, $b_1=b_2=\sqrt{2}a$, $\gamma=45^{\circ}$; (ii) within the hexagonal SkL, $b_1=a=b_2/\sqrt{3}$, $\gamma=30^{\circ}$.

\begin{figure*}
\includegraphics[width=1.99\columnwidth]{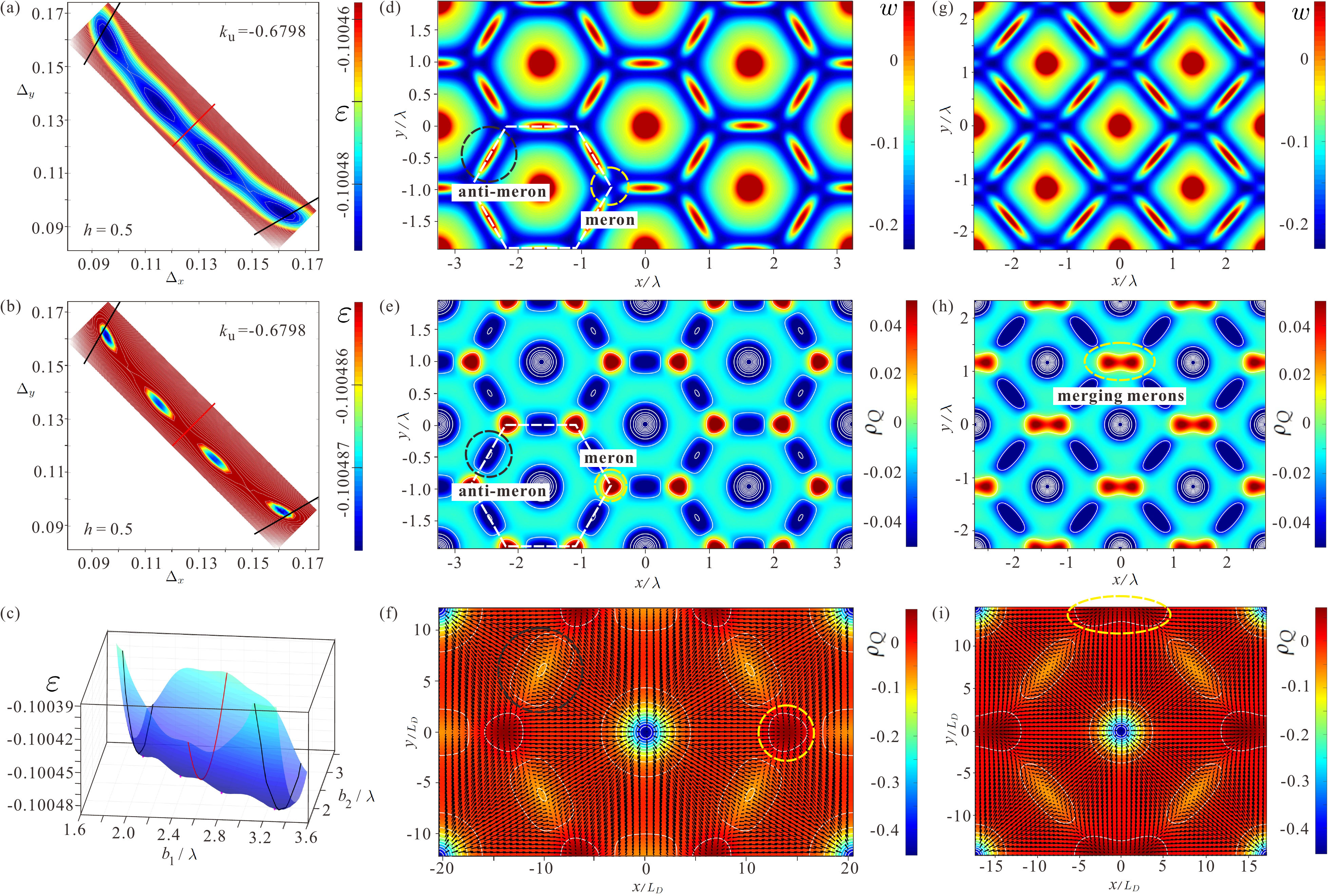}
\caption{(color online) (a) The color plot of the energy density $\varepsilon$ on the plane $\Delta_x\Delta_y$. Black and red lines show the parameters
for the hexagonal and square SkLs, correspondingly. The legend highlights the energy range $\varepsilon \in [\varepsilon_{min},\varepsilon_{min}+3 \times 10^{-5}]$. 
(b) The same energy plot, but in the range $\varepsilon \in [\varepsilon_{min},\varepsilon_{min}+3 \times 10^{-6}]$ to prove the same depth of all four energy minima.
(c) The energy density of distorted skyrmion orders plotted as a surface.  The hexagonal SkLs (highlighted by the black curves) almost reach global energy minima.
The parameters for the square SkL constitute a red curve with the minimum being a saddle point, i.e., there is no solution for
the square SkL for this value of the EPA. Pink dots show the actual energy minima of the system.
(d), (g) Contour plots for the energy density distributions $w(x,y)$ within the hexagonal and the distorted SkLs, correspondingly, for $h = 0.5,\, k_u = -0.6798$. In both graphs, the legends highlight the same energy range $[w_{min},w_{min}+0.3]$. (e), (h) Color plots of the topological charge density $\rho_Q$ in both skyrmion arrangements. The legends "zoom" the interval $[-0.05, 0.05]$. (f), (i) Distributions $\rho_Q(x,y)$ with the legends exhibiting the intervals $[\rho_{Q}^{min}, \rho_{Q}^{max}]$. Black arrows show the projections of the magnetization vectors onto the plane $xy$. 
\label{fig05}}
\end{figure*}

As an example, Fig. \ref{fig02} (c) shows the color plot of the energy density depending on the cell sizes for $h=0.5, \,k_u=0.6785$. The red and black lines highlight the grid spacings for the square ($\Delta_x=\Delta_y$) and the hexagonal ($\Delta_x=\Delta_y\sqrt{3}$) skyrmion lattices. The energy minimum corresponds to a slightly distorted SkL with the lattice parameters $\Delta_x^{min}=0.161,\, \Delta_y^{min}=0.095$, which is very close to the hexagonal skyrmion ordering and  corresponds to $b_1=N_x\Delta_x^{min}/4\pi=3.28,\,b_2=1.94$ and $\gamma=30.54^{\circ}$. The other energy minimum is reached for the interchanged lattice parameters $\Delta_y^{min}=0.161,\, \Delta_x^{min}=0.095$ when the energy contour plot is mirrored with respect to the red line. In the following, we will consider only the lower part of the total energy density distribution (compare with the color plot in Fig. \ref{fig05} (g), (h)).

The energy density in Fig. \ref{fig02} (c) is computed as follows:
\begin{equation}
\varepsilon = (1/V)\int w(x,y)dV, \,V=N_xN_yN_z\Delta_x\Delta_y\Delta_z, \nonumber
\end{equation}
where $V$ is the volume of the unit cell in Fig. \ref{fig02} (a). $N_x=256, N_y=256, N_z=1$. 
To highlight the topology of the energy surface in the direct  vicinity of the energy minimum, the color plot discerns the energy range
from the minimal energy value to $\varepsilon_{min}+3 \times 10^{-5}$.

\section{Reorientation transition between hexagonal and square skyrmion arrangements \label{sect03}} 

Fig. \ref{fig03} shows a series of energy "fingerprints" for the increasing value of the negative EPA $k_u$. A new minimum corresponding to a distorted SkL is clearly visible to form in (a) for $k_u=-0.6790$. It gradually deepens  (Fig. \ref{fig03} (b), (c)) and equals  the energy minimum of the hexagonal SkL at $k_u^0=-0.6798$. After this anisotropy value, the hexagonal SkL becomes a metastable state. At the value $k_u=-0.6817$ the  local energy minimum of the hexagonal SkL disappears, and the global minimum corresponds to a square lattice, which becomes fully shaped at $k_u=-0.6826$. Thus, in the anisotropy range $k_u\in [-0.6817,-0.6788]$,  there are coexisting solutions for two SkLs, which underlie the first-order phase transition. 

The lattice parameters for both skyrmion orders are shown in Fig. \ref{fig04} (a). The red (blue) solid lines correspond to the hexagonal (distorted) SkL, which is the global minimum of the system whereas the dotted lines -- to metastable solutions. The dashed vertical lines indicate the hysteresis behavior of the reorientation transition and highlight the limiting anisotropy values of the loop as well as the critical value of the anisotropy $k_u^0$. The dark-blue line corresponds to the square SkL with equal parameters, $b_1=b_2$. The angle $\gamma$ (Fig. \ref{fig04} (b)) changes almost linearly until it reaches the value $\gamma \approx 45^{\circ}$ when the anisotropy value $k_u^0$ is surpassed and stays almost intact below this point ($\gamma \approx 30^{\circ}$). 

The underlying reason of this phase transition can be elucidated from the energy density distributions $w(x,y)$ as well as topological charge densities $\rho_Q$ within different skyrmion arrangements. Fig. \ref{fig05} features hexagonal and distorted SkLs for $k_u^0=-0.6798$ when corresponding energy minima are equal. 

In Fig. \ref{fig05} (a), (b), we plot the color plots for the energy density $\varepsilon$ to show all four solutions: solutions above the red line are rotated by $90^{\circ}$ with respect to the solutions below the red line. Fig. \ref{fig05} (b) zooms the energy landscape in the direct vicinity of the energy minima, $\varepsilon \in [\varepsilon_{min},\varepsilon_{min}+3 \times 10^{-6}]$. We also remark that these color plots are just top views of the 3D energy density surfaces (Fig. \ref{fig05} (c)).

\begin{figure*}
\includegraphics[width=1.9\columnwidth]{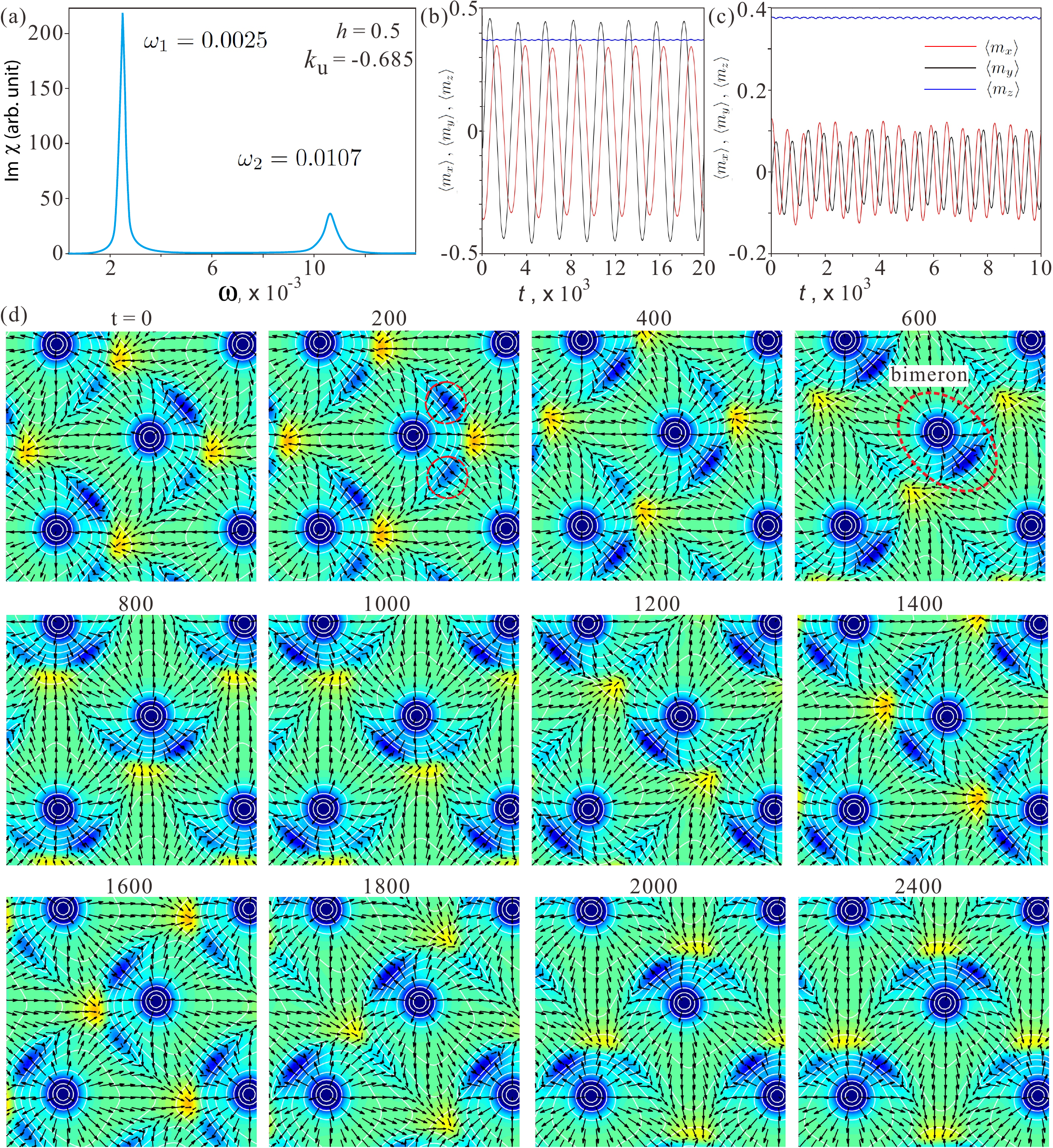}
\caption{(color online) (a) Imaginary part of the  in-plane dynamical susceptibility for $k_u=-0.685,\, h=0.5$ exhibiting two resonance frequencies. (b), (c) Calculated time evolutions of the averaged magnetization components  $\left\langle m_x\right\rangle, \,\left\langle m_y\right\rangle,\,\left\langle m_z\right\rangle$ in both spin-wave modes. (d) Spin dynamics within the low-frequency mode characterized by the snapshots of the topological charge density (see text for details). 
\label{fig07}}
\end{figure*}

For the relatively large value of the easy-plane anisotropy $k_u^0$, the role of merons formed within the domain boundaries between skyrmion cells becomes paramount as will be seen later.  According to Ref. \cite{Lin}, merons emerge due to the overlap of neighboring skyrmions.

Let's first scrutinize the internal structure of  the hexagonal SkLs.
The anti-merons (highlighted by dashed black circles in Figs. \ref{fig05} (d) - (f)) have the positive energy density (Fig. \ref{fig05} (d)) and the negative topological charge density (Fig. \ref{fig05} (e), (f)). They also have the negative vorticity (Fig. \ref{fig05} (f)). Both figures \ref{fig05} (e) and \ref{fig05} (f) show the topological charge density although Fig. \ref{fig05} (e) zooms more of the  interval $\rho_Q \in [-0.05,0.05]$. Fig. \ref{fig05} (f) shows just two unit cells with the in-plane components of the magnetization vectors as black arrows. 
Since each anti-meron is shared by two adjacent skyrmion cells (highlighted by dashed white hexagons), there are three anti-merons per unit cell. The central vortex also represents an anti-meron. 
Although it has the positive vorticity, its negative polarity endows it with the negative topological charge. 

Merons (highlighted by dashed yellow circles in Figs. \ref{fig05} (d) - (f)) exhibit the positive topological charge density (Fig. \ref{fig05} (e), (f)) and the negative energy density (Fig. \ref{fig05} (d)), which is the outcome of their positive vorticity and polarity.
 Since merons are located in the corners, they are shared by three neighboring unit cells, which amounts to two merons per unit cell.

In the distorted SkL (Figs. \ref{fig05} (g) - (i)), two corner merons approach each other and eliminate one anti-meron located in-between.  Still, two remaining merons are well-discernible (as highlighted by dashed yellow ellipses in (h), (i)). The energetic advantage of the distorted SkL due to the collapse of anti-merons is, however, counterbalanced by slightly higher skyrmion density. At larger anisotropy values, two merons merge into one and the lattice becomes a perfect square arrangement of skyrmions as shown in Fig. \ref{fig04} (c) (last panel). The unit cell includes one meron and two anti-merons. 

The results of the current section imply that, at high anisotropy values, the internal structure of the SkL can be considered from the point of view of interacting merons confined within the stretched domain boundaries (Fig. \ref{fig04} (c)). Such a network of merons was dubbed a square vortex-antivortex crystal in Ref. \cite{Lin}. Alternatively, one can call it "a square meron-antimeron crystal".

\section{Spin-Wave Modes of the square SkL \label{sect05}} 

It is instructive to investigate the dynamics of the meron-antimeron crystal under external oscillating magnetic fields and to deduce whether it becomes different as compared with the excitation effects in hexagonal SkLs \cite{Mochizuki}. 

We will study collective spin dynamics of meron crystals for $k_u=-0.685,\, h=0.5$ following the numerical procedure explicitly described in Ref. \cite{Mochizuki}.

First of all, we adapt the discretized version of equation (\ref{functional}) with the same energy terms: 
\begin{align}
&w(\mathbf{S}) =  J\,\sum_{<i,j>} (\mathbf{S}_i \cdot \mathbf{S}_j ) -\sum_{i} \mathbf{H} \cdot \mathbf{S}_i  - K_u (\mathbf{S}_i\cdot \hat{z})^2\nonumber\\
&- D \, \sum_{i} (\mathbf{S}_i \times \mathbf{S}_{i+\hat{x}} \cdot \hat{y} - \mathbf{S}_i \times \mathbf{S}_{i+\hat{y}} \cdot \hat{x}). 
 \label{model}
\end{align}
We consider classical spins $\mathbf{S}_i$ of unit length on a square two-dimensional lattice, where $<i,j>$ denote pairs of nearest-neighbor spins.
All calculations are performed for spin systems with $104\times 104$ sites under pbc, which include four unit cells. The spin configuration was minimized with respect to the period and  corresponds to the energy minimum for the chosen control parameters. 
%
%
The DMI constant $D = J \tan (2\pi/\lambda)$ defines the period of modulated structures $\lambda$ (\ref{lambda}).
In what follows, we use $J=-1$ and the DMI constant is set to $0.3249$, i.e., $\lambda\approx 20$.

Next, we solve numerically the LLG equation under time-dependent ac magnetic fields. We use the fourth-order Runge-Kutta method.
The equation is given by 
\begin{equation}
\frac{\partial\mathbf{S}_i}{\partial t}=-\frac{1}{1+\alpha_G^2}[\mathbf{S}_i\times\mathbf{H}^{eff}_{i}+\frac{\alpha_G}{S}\mathbf{S}_i\times(\mathbf{S}_i\times\mathbf{H}^{eff}_{i})],
\end{equation}
where $\alpha_G$ is the dimensionless Gilbert-damping coefficient. We used a rather small dimensionless damping parameter, $\alpha_G=0.01$, to make visible all  peaks in the imaginary part of dynamical magnetic susceptibility (Fig. \ref{fig07} (a)). 
$\mathbf{H}^{eff}_{i}$ is a local effective field acting on the \textit{i}th spin $\mathbf{S}_i$ and derived from the Hamiltonian $\mathbf{H}^{eff}_{i}=-\partial w/\partial\mathbf{S}_i$.

To study the microwave-absorption spectra due to spin-wave resonances in the square SkL, we apply in-plane  $\delta$-function pulses of magnetic field $h^{\omega}=0.1$ at $t=0$ and then trace spin dynamics. 
The absorption spectrum of the imaginary part of the dynamical susceptibility, $\mathrm{Im}\chi(\omega)$ is calculated from the Fourier transformation of the magnetization $\mathbf{m}=(1/N)\sum\mathbf{S}_i(t)$.
Fig. \ref{fig07} (a) shows  the imaginary part of the in-plane dynamical magnetic susceptibility in dependence on $\omega$ for the chosen control parameters. 
The calculated spectrum for $h^{\omega}$ parallel to the $y$ axis exhibits two resonance peaks at $\omega_1=0.0025$ and $\omega_2=0.0107$.

To identify each spin-wave mode, we trace the spin dynamics by applying an oscillating magnetic field with a corresponding resonant frequency and the amplitude $h^{\omega}=0.001$ (see supplementary videos).  Figs. \ref{fig07} (b), (c) show average components of the magnetization in each mode, $\left\langle m_x\right\rangle, \,\left\langle m_y\right\rangle,\,\left\langle m_z\right\rangle$. In Fig. \ref{fig07} (d),  we display calculated time evolutions of the spins in the first mode. The in-plane projections of the spins are represented by black arrows; the color plots  are the topological charge distributions zoomed in the interval $\rho_Q \in [-0.05,0.05]$. 

The high-frequency mode represents a CCW rotation of the central anti-meron as well as the square network of domain boundaries (see supplementary videos). This mode is analogues to the CCW rotation of the hexagonal SkL found in Ref. \cite{Mochizuki}. Although the supplementary video gives a particular emphasis to the DW-merons, still, the topological charge density concentrates around the center of the unit cell. Interestingly, the "intensity" of the topological charge within DW anti-merons variates depending on the position of the central anti-meron within the square unit cell. The amplitude of the ac field is small enough to exclude any coupling among merons. 

The low-frequency mode, however, is hard to anticipate. Virtually, the central anti-merons remain immobile (Fig. \ref{fig07} (d)). It is a network of domain walls, which rotate and come into contact with the central anti-meron in turn with each side of the unit cell. In the point of contact, the structure of the DW anti-meron becomes pronounced and accumulates large topological charge density (see, e.g., the snapshot in Fig. \ref{fig07} (d) for $t=600$). Other DW merons are barely discernible. Their topological charges are small as compared with the topological charge $Q=-1$ of a formed bimeron (encircled by the dashed red line).
When the central anti-meron moves to the next corner (or rather the square matrix rotates) and again creates a bimeron structure but now with another side of the unit cell, the topological charge is first split between two DW anti-merons (e.g., for $t=200$; such anti-merons are encircled by the dashed red lines) when the central anti-meron is located in the corner of the unit cell. Thus, the low-frequency rotational mode is based on the creation and annihilation of DW merons and enables coupling between anti-merons. 

\begin{figure}
\includegraphics[width=0.99\columnwidth]{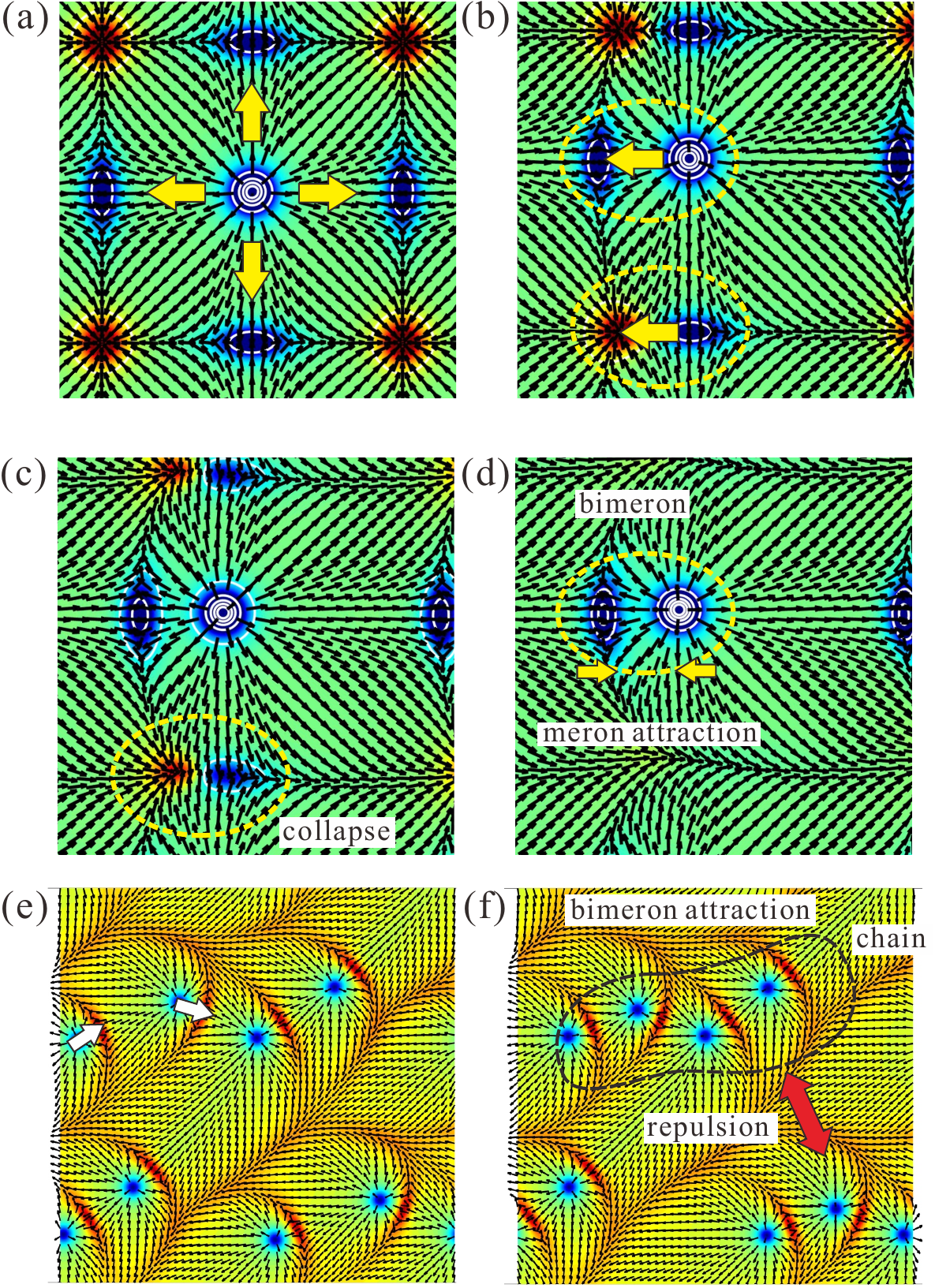}
\caption{(color online) Mutual transformation between the square SkL and the elliptical cone (tilted FM state) occurring via collapse and coupling of merons. The internal structure of states in (a) - (d) is shown as color plots of the topological charge density. The color plots in (e), (f) exhibit the $m_z$-component of the magnetization. The elliptical cone in these figures contains a network of attracting bimerons (see text for details). 
\label{fig06}}
\end{figure}

The considered rotational process also gives a hint at the possible scenario of the first-order phase transition between the square meron-antimeron crystals and tilted FM states or elliptical cones.

\section{The first-order phase transition between 2D square SkLs and 1D elliptical cones \label{sect04}} 

With the increasing EPA, the period of the square SkL gradually increases and diverges at $k_u\approx -0.765$.
Since the inter-meron distances also increase, this results in the excessive energy of this phase.  
The first-order phase transition with the elliptical cone is computed to take place at somewhat lower EPA, $k_u=-0.742$. 
%

The first step of such a transition is to create some asymmetry in the balanced position of the central anti-meron when it is attracted simultaneously by four anti-merons in the domain walls (Fig. \ref{fig06} (a)). In Fig. \ref{fig06} (b), the central anti-meron is shifted to the left and is thus attracted by the corresponding boundary anti-meron to reach the minimum of their interaction potential \cite{Mukai24}. 

At the same time, domain-wall merons and anti-merons approach each other what creates prerequisites for their mutual annihilation (Fig. \ref{fig06} (c)). Notice that the domain boundaries bend when the anti-meron acquires the crescent shape typical for bimerons; this additionally facilitates the collapse of DW merons. 

After the collapse of excessive merons, the remaining bimerons additionally adjust the value of their  dipole moments to reach the minimum of the inter-meron potential 
 (Fig. \ref{fig06} (d)). 

Remarkably, in different square unit cells, the central anti-merons may be drawn randomly by either side. As a result, the dipole moments of the formed bimerons may form complex bimeron tesselations, so called bimeron polymers \cite{Mukai24}. Since bimerons also attract each other, they may locally assemble into chains or looped clusters dubbed "roundabouts" and "crossings" in Ref. \cite{Mukai24}. Fig. \ref{fig06} (e) shows bimerons with mutually perpendicular dipole moments (white arrows). In Fig. \ref{fig06} (f), bimerons attempt to align into chains with the parallel orientation of dipoles. Neighboring chains, on the contrary, repel each other. The regions between different bimeron clusters are filled by the elliptical cone or the tilted FM state.  In fact, the internal structure of a cone is quite close to the tilted FM state for the chosen control parameters. In this sense, the final state can be viewed as a homogeneous state accommodating isolated bimerons and bimeron clusters.

\section{Conclusions}

In the present paper, I focused on the essential role of merons arising within domain boundaries between skyrmion cells in chiral magnets with the easy-plane anisotropy. Being barely noticeable at the spin distributions and possessing small topological charges, as compared with skyrmions, they nevertheless (i) act as drivers of the structural phase transition between hexagonal and square skyrmion lattices as well as of the first-order phase transition between SkLs and tilted FM states; (ii) define the dynamic properties of square meron-antimeron crystals. 

In particular, I show that merons located in the corners of the hexagonal unit cells merge and thus "erase" anti-merons with the positive energy density located in-between. This process triggers the structural phase transition from the hexagonal to square skyrmion order. Mutual annihilation of merons and anti-merons underlies the subsequent phase transition into the homogeneous state. 
Anti-merons with the opposite polarities are shown to couple and form bimerons, which, on the higher level, gather into bimeron networks. 

Interestingly, the coupling of anti-merons defines the character of the collective modes induced by the oscillating in-plane fields. In the low-frequency mode, the square-shaped array of domain walls circles around the central anti-meron and lets it form a bimeron state with each DW anti-meron successively. During this process, the coupled DW anti-meron develops the crescent shape and accumulates the topological charge whereas the DW anti-merons within other sides of the unit cell almost decay. In the high-frequency mode, no creation or annihilation of merons is observed; the merons undergo rotations as would be expected for conventional skyrmions. 

I argue that the non-trivial findings of the present paper complement previous studies from both fundamental and applied points of view and consider the role of merons from different perspective.

\textit{Acknowledgements. }
The author is grateful to Natsuki Mukai, Kaito Nakamura and Takayuki Shigenaga for useful discussions. 

\end{document}